\documentclass[preprint2]{aastex}
\usepackage[colorlinks,urlcolor=blue,citecolor=blue,linkcolor=blue]{hyperref}

\shorttitle{Large-mass neutron stars with hyperonization}
\shortauthors{Jiang et al.}
\begin{document}

\title{Large-mass neutron stars with hyperonization}

\author{Wei-Zhou Jiang}
\affil{ Department of Physics, Southeast University, Nanjing 211189,
China\\
and Center of Theoretical Nuclear Physics,
National Laboratory of Heavy Ion Accelerator, Lanzhou 730000, China}

\author{Bao-An Li}
\affil{Department of Physics and Astronomy,
Texas A\&M University-Commerce, Commerce, TX 75429, USA\\
and Department of Applied Physics, Xian Jiao Tong University, Xian 710049, China}

\and
 \author{Lie-Wen Chen}
\affil{ INPAC, Department of Physics and Shanghai Key Laboratory for
Particle Physics and Cosmology, Shanghai Jiao Tong University,
Shanghai 200240, China\\
and Center of Theoretical Nuclear Physics,
National Laboratory of Heavy Ion Accelerator, Lanzhou 730000, China}

\begin{abstract}

Within a density-dependent relativistic mean-field model using
in-medium meson-hadron coupling constants and meson masses, we
explore effects of in-medium hyperon interactions on properties of
neutron stars. It is found that the hyperonic constituents in
large-mass neutron stars can not be simply ruled out, while the
recently measured mass of the millisecond pulsar J1614-2230 can
constrain significantly the in-medium hyperon interactions. Moreover,
effects of nuclear symmetry energy on hyperonization in neutron stars
are also discussed.
\end{abstract}

\section{Introduction}

The in-medium hyperon interactions play an important role in determining
properties of hypernuclei and the hyperonization in neutron stars.
Conversely, observed properties of hypernuclei and neutron stars can be used to constrain
the in-medium hyperon interactions. For instance, it has been shown that
the properties of hypernuclei are indeed very useful in
extracting in-medium hyperon potentials at subsaturation
densities~\citep{Fr2007}. In bulk matter, such as neutron stars,
hyperons can be produced by virtue of strong interactions. They can
actually become important constituents of neutron stars and thus have
important effects on astrophysical observations (for a review,
see~\citep{Gl2001}). In fact, it is well known that the
hyperonization can reduce the maximum mass of neutron stars as much
as 3/4$M_\odot$~\citep{Gl1985,Gl1991,Ji2006}. Currently, a number of
phenomenological models, considering only the minimum compositions of
nucleons and leptons using interaction parameters that are well
calibrated by using terrestrial nuclear laboratory data, can produce
the maximum mass of neutron stars around or below 2$M_\odot$
~\citep{da02,Pi2007}. Interestingly, several neutron stars with large
masses around $2M_\odot$ have been observed~\citep{Ni05,Ni08,Oz06}
recently. In particular, the $2M_\odot$ pulsar J1614-2230 was
measured rather accurately through the Shapiro delay~\citep{De10}.
Since properties of neutron stars are determined by the nuclear
equation of state (EOS) and the hyperonization reduces significantly
the maximum mass of neutron stars, it has been stated that these
observations seem to rule out almost all currently proposed hyperon
EOS. Recent evidence for this can also be found in the work of the
Brueckner approach~\citep{Sch2011}.

Though most of the hyperon EOSs give lighter neutron stars, there
were actually a few endeavors in the past to stiffen the EOS either
by invoking strong repulsions for hyperons or pushing upwards the
onset density of hyperons, leading to heavy neutron stars involving
hyperons~\citep{Ho01,Ta02,St07,De08}. Recently, by virtue of
nonlinear self-interacting terms involving a vector meson with hidden
strangeness, Bednarek et al. obtained a stiff hyperon EOS with which
the large mass of the PSR J1614-2230 can be produced~\citep{Be2011}.
Usually, the SU(6) relations are imposed to constrain the
meson-hyperon coupling constants. Such SU(6) relations were recently
reexamined by Weissenborn et al. and an arbitrary breaking of such
relations can also result in the stiffening of hyperon EOS and the
uplift of the maximum mass of neutron stars with the inclusion of
hyperons~\citep{We11}. On the other hand, the quark deconfinement may
occur in the medium as the spatial overlap of nucleons becomes
sufficient to dissolve the boundary of color singlets with the
increase of density. Of course, the quantitative understanding of
such a color deconfinement in cold medium is still model dependent.
Typical effective QCD models include the NJL-like models, e.g.,
see~\citep{Kl2007,Ip2008,Pa2008,Bo2012} and widely used MIT bag
models, e.g., see~\citep{Pr1997,Al05,We11b}, as well as
Schwinger-Dyson approaches~\citep{Li2011}. With the Maxwell or Gibbs
constructions for the hadron-quark phase coexistence, the resulting
quark EOS may give rise to two possible types of stars: strange stars
which are totally made of absolutely stable strange
matter~\citep{Gl2000} and hybrid stars with a quark core and hadron
out-layer. In order to be consistent with the recent observation of
the 2$M_\odot$ pulsar, the strong coupling and/or color
superconductivity were shown to be
necessary~\citep{Xu03,Al05,Kl2007,Ip2008,Pa2008,We11b,Bo2012}. While
the compositions of hybrid stars are rather model dependent, it is
interesting to mention that Yasutake et al. suggested the hyperon
suppression with the MIT model using a density-dependent bag
constant~\citep{Ya2011}.

Noticing the recent investigations on the consequences of various
quark EOS, we examine in this work the consistency of the hyperon EOS
with the recent observation of the PSR J1614-2230. Despite  the
impressive progress made in recent decades in constraining the
nuclear EOS using both astrophysical observations and nuclear
reaction data, see, e.g.~\citep{You99,da02,Li08}, many uncertainties
still remain. The in-medium hyperon interactions are among the most
uncertain ingredients of neutron star models. We shall thus seek
in-medium hyperon interactions that can produce the observed maximum
mass of neutron stars. In view of the fact that some microscopic
theories, such as the Brueckner approach, are still having
difficulties to obtain the $2M_\odot$ of hyperonized neutron stars,
here we resort to the phenomenological models developed in
refs.~\citep{Ji2007,Ji2007b} to analyze effects of various in-medium
hyperon interactions on properties of neutron stars.

\section{Density-dependent relativistic mean-field models and parametrizations}
In order to study conveniently the in-medium interactions for
hyperons, we seek for density-dependent relativistic models without
nonlinear interactions. In our previous works~\citep{Ji2007,Ji2007b},
we constructed density dependent relativistic mean-field (RMF) models
using in-medium hadron properties according to the Brown-Rho scaling
due to the chiral symmetry restoration at high
densities\citep{br91,song01,br05,br07}. In these models, the
symmetric part of the resulting equations of state around normal
density is consistent with the data of nuclear giant monopole
resonances~\citep{You99} and at supra-normal densities it is
constrained by the collective flow data from high energy heavy-ion
reactions~\citep{da02}, while the resulting density dependence of the
symmetry energy at sub-saturation densities agrees with that
extracted from the isospin diffusion data from intermediate energy
heavy-ion reactions\citep{ts04,ch05,li05}. Our models with the chiral
limits are soft at intermediate densities but stiff at high densities
naturally, producing a heavy maximum neutron star mass around
2$M_\odot$. It is interesting to see that the EOS extracted from the
celestial observations most recently features similar
characters~\citep{St2010}. Apart from the usual studies on the
minimum constituents in neutron stars with electrons, protons and
neutrons, in this work we include the hyperonic degrees of freedom to
study the in-medium interactions for hyperons with the constraints of
the recent celestial observations. The model Lagrangian with the
density-dependent parameters is written as
\begin{eqnarray}
 {\cal L}&=&
{\overline\psi}_B[i\gamma_{\mu}\partial^{\mu}-M_B^* +g^*_{\sigma
B}\sigma-g^*_{\omega
B} \gamma_{\mu}\omega^{\mu}\nonumber\\
&& -g^*_{\rho B}\gamma_\mu \tau_3 b_0^\mu]\psi_B
 +\frac{1}{2}(\partial_{\mu}\sigma\partial^{\mu}\sigma-m_{\sigma}^{*2}\sigma^{2})
 \nonumber\\
 && - \frac{1}{4}F_{\mu\nu}F^{\mu\nu}+
\frac{1}{2}m_{\omega}^{*2}\omega_{\mu}\omega^{\mu}
-\frac{1}{4}B_{\mu\nu} B^{\mu\nu} \nonumber\\
 &&     +
\frac{1}{2}m_{\rho}^{*2} b_{0\mu} b_0^{\mu}  +{\cal L}_Y +{\cal L}_l,
\label{eq:lag1}
\end{eqnarray}
 where $\psi_B,\sigma,\omega$, and $b_0$ are the fields
of the baryons, scalar, vector, and isovector-vector mesons, with
their  masses $M_B^*, m^*_\sigma,m^*_\omega$, and $m^*_\rho$,
respectively. $F_{\mu\nu}$ and $ B_{\mu\nu}$  are the strength
tensors of the $\omega$ and $\rho$ mesons, respectively. The meson
coupling constants and masses with asterisks denote the density
dependence, given by the BR scaling~\citep{Ji2007,Ji2007b}. ${\cal
L}_l$ and ${\cal L}_Y$ are the Lagrangian for leptons and hyperons,
respectively. The parameters for strange mesons $\sigma^*$ (i.e.
$f_0$, 975MeV) and $\phi$ (1020MeV) in ${\cal L}_Y$  are assumed to
be density independent.

The energy density and pressure in the RMF approximation read,
respectively,
\begin{eqnarray}
{\mathcal{E}}&=&\frac{1}{2} m_\omega^{*2} \omega_0^2 +\frac{1}{2}
m_\rho^{*2} b_{0}^2 +\frac{1}{2} m_\phi^{2} \phi_0^2+
 \frac{1}{2} m_\sigma^{*2}\sigma^2 \nonumber\\
 && + \frac{1}{2} m_{\sigma^*}^{2}{\sigma^*}^2 +\sum_{i}
 \frac{2}{(2\pi)^3}\int_{0}^{{k_F}_i}\! d^3\!k~ E_i^*,\\
  \label{eqe1}
 p&=&\frac{1}{2} m_\omega^{*2} \omega_0^2
+\frac{1}{2} m_\rho^{*2} b_{0}^2+ \frac{1}{2} m_\phi^{2} \phi_0^2\nonumber\\
&& -\frac{1}{2} m_\sigma^{*2}\sigma^2 -\frac{1}{2}
m_{\sigma^*}^{2}{\sigma^*}^2
   -\Sigma^R_0\rho \nonumber\\
 && +  \frac{1}{3}\sum_{i}\frac{2}{(2\pi)^3}\int_{0}^{{k_F}_i}\! d^3\!k
 ~\frac{{\bf k}^2}{E^*_i},
\label{eqp1}
\end{eqnarray}
where $i$ runs over the species of baryons and leptons considered in
neutron star matter, $E^*_i=\sqrt{{\bf k}^2+m_i^{*2}}$ with $m_i^*$
being the Fermion effective mass of species $i$, and $\Sigma^{R}_0$
is the rearrangement term, originating from the density-dependent
parameters, to preserve the thermodynamic consistency~\citep{Ji2007}.

The density dependence of parameters is described by the scaling
functions  that are the ratios of the in-medium parameters to those
in the free space. For the nucleonic sector, we take the scaling
functions for the coupling constants of scalar and vector mesons as
$\Phi_{\sigma N}(\rho)={(1-y_i\rho/\rho_0)}/{(1+x_i\rho/\rho_0)}$
with coefficients $x_i$ and $y_i$ given in~\citep{Ji2007,Ji2007b}.
For the hyperonic sector, we will give the scaling form for coupling
constants below. For hadron masses, the scaling function is given as
$\Phi(\rho)=1-y\rho/\rho_0$.

For the neutron star matter with hyperonizations, the chemical
equilibrium is established on the weak interactions of baryons and
leptons. We study  chemically equilibrated and charge neutral matter
including baryons $(N,\Lambda,\Sigma,\Xi)$ and leptons $(e,\mu)$.
Note that although the baryon chemical potential is modified by the
rearrangement term, the chemical equilibrium is independent of this
modification.

In the present work, the RMF  parameter sets SLC and
SLCd~\citep{Ji2007b} that have  no mass scalings for baryons are
extended to include the hyperonizations. The main difference between
the SLC and SLCd is their prediction for the density dependence of
nuclear symmetry energy. More quantitatively, the slope parameter $L$
at saturation density is $L=92.3$ MeV and $61.5$ MeV for the SLC and
SLCd, respectively. Assuming all analyses of terrestrial data are
equally reliable, a conservative estimate puts $L$ in the range of
approximately 25 to 115 MeV and the symmetry energy at normal density
$E_{sym}(\rho_0)$ to be between 26 and 34 MeV~\citep{Ne2011,ts12}.
However, it is interesting to note that the majority of the analyses
of terrestrial nuclear experiments scatter the $L$ and
$E_{sym}(\rho_0)$ values around 60 MeV and 30 MeV, respectively.
Several recent analyses of astrophysical observations and phenomena
including the mass-radius correlation~\citep{St2012a}, the binding
energy of neutron stars~\citep{Newton09}, the frequencies of
torsional crustal vibrations~\citep{St09,Gear11} and the $r$-model
instability window~\citep{Wen2012}, all consistently favor $L$ values
less than about 70 MeV. For instance, the latest analysis of neutron
star observation puts $L$ in a somewhat lower range of 36 to 55 MeV
at 95\% confidence level~\citep{St2012a}. It is worth noting that the
$L$ value from the SLCd model is more consistent with the most
stringent constraint on $L$ so far, i.e. approximately $40 < L < 60$
MeV, obtained by combining constraints from analyzing both
terrestrial experiments and astrophysical
observations~\citep{Lat2012}. By comparing calculations with the SLC
and SCLd parameter sets we shall examine effects of the symmetry
energy on the mass-radius correlation of neutron stars, though  the
SLCd model can describe the available observations more consistent
with conclusions based on other analyses in the literature. The
parametrization for hyperons is elaborated in the following. The
coupling of mesons with hyperons can generally be given in terms of
the parameters $X_{\sigma Y}$, $X_{\omega Y}$, and $X_{\rho Y}$,
which are ratios of the meson coupling with hyperons to that with
nucleons. Lack of strong constraints on these
parameters~\citep{We11},  a variety of choices of these parameters
roughly varying from 0.2 to unity were used in practical
studies~\citep{Gl1991,Av2006}. Considering that the nucleonic sector
of our models respects chiral limits at high densities, we assume for
hyperons two cases:  the usual case (UC) that the hyperons have a
similar medium effect to nucleons and the separable case (SC) that
the meson-hyperon coupling constant is separated into
density-dependent and density-independent parts regardless of the
chiral limit constraint on the strange sector in hyperons.
Nevertheless, the hyperon potentials~\citep{Mi1988,Ha1989,Fu1998}
\begin{equation}\label{eqhpot}
U^{(N)}_\Lambda=-30 MeV=-U^{(N)}_\Sigma, \hbox{ } U^{(N)}_\Xi=-18
MeV,
\end{equation}
in nuclear matter at saturation density are used to preserve the
relation  between the vector and scalar meson coupling constants.
Note that the repulsive $\Sigma$ hyperon potential is invoked
here~\citep{Ma1995,No2002}. For the strange mesons, we adopt the
density-independent coupling constants in both cases for simplicity.
The potentials for the $\Lambda$ and $\Xi$ hyperons in $\Xi$ matter
$U^{(\Xi)}_\Lambda=U^{(\Xi)}_\Sigma=U^{(\Xi)}_\Xi=-40$ MeV are used
to obtain the coupling constants of the strange mesons (For the
details, see~\citep{Ji2006,Sc2000}).

\begin{figure}[thb]
\epsscale{1.0} \plotone{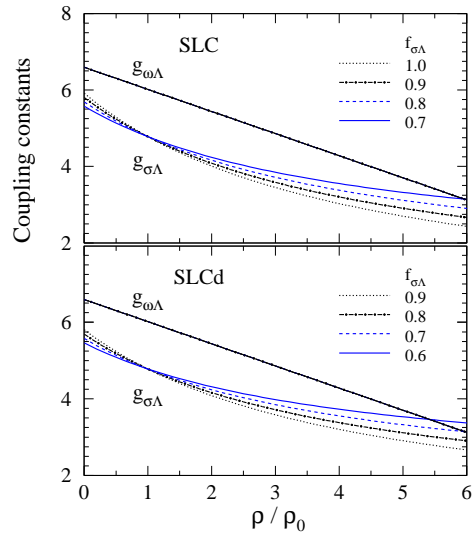}\caption{(Color online) Coupling
constants of the $\sigma$ and $\omega$ mesons with the $\Lambda$
hyperon as a function of density. \label{fcoup}}
\end{figure}
The ratio parameters for coupling constants in the UC are assumed to
be  density-independent. In the SC, we consider following scaling
functions for the meson-hyperon coupling constants that consist of
two terms:
\begin{eqnarray}\label{eqphiY}
&&\Phi_{\omega \Lambda}(\rho)=\Phi_{\omega
\Sigma}(\rho)=\frac{1}{3}\Phi_{\omega N}(\rho_0)
+\frac{2}{3}\Phi_{\omega N}(\rho),\nonumber\\
&&\Phi_{\omega \Xi}(\rho)=\frac{2}{3}\Phi_{\omega N}(\rho_0)
+\frac{1}{3}\Phi_{\omega N}(\rho),\\
&&\Phi_{\sigma Y}(\rho)=(1-f_{\sigma Y})\Phi_{\sigma N}(\rho_0)
+f_{\sigma Y}\Phi_{\sigma N}(\rho),\nonumber
\end{eqnarray}
where $\rho_0$ is the saturation density and  $f_{\sigma Y}$ is an
adjustable constant. The scaling functions $\Phi_{\rho\Sigma}$ and
$\Phi_{\rho\Xi}$ for the $\rho$ meson are taken the same as those
of the $\omega$ meson. Note that $\Phi_{\omega N}(\rho_0)$ and
$\Phi_{\sigma N}(\rho_0)$ are just constants.  The factors before
$\Phi(\rho_0)$ and $\Phi(\rho)$ implied from constituent quark
compositions play a role in averaging on the hadron level the
coupling constant between the density-dependent part originating from
the chiral limit and the density-independent part that is presumably
attributed to the strange sector in hyperons. The form in
(\ref{eqphiY}) produces the relation $\Phi_{iY}\equiv\Phi_{iN}$ at
saturation density. Thus, we do not need to readjust the parameter
$g_{\sigma Y}(\rho_0)$ as $f_{\sigma Y}$ changes.  For a few choices
of $f_{\sigma \Lambda}$, we plot the meson-$\Lambda$ hyperon coupling
constants  in Fig.~\ref{fcoup}. It is seen that the larger the
$f_{\sigma Y}$, the smaller the $g_{\sigma Y}$ at high densities.

  \begin{table*}[ht]
  \begin{center}
\caption{Meson-hyperon coupling constants in various cases with
models SLC and SLCd. In the UC, we tabulate the values with
$X_{\omega \Lambda}=X_{\omega \Sigma}=X_{\omega \Xi}=0.9$. The
coupling constants in the medium are obtained as $g_{i Y}^*=g_{i
Y}^0\Phi_{iY}(\rho)$. In the UC, $g_{i Y}^0=g_{iY}(\rho=0)$, while in
the SC they are not equal. In the SC, the parameters listed here are
free of the parameter $f_{\sigma Y}$. For the couplings with the
$\Sigma$ hyperon, we take in the calculation the relations:
$g_{\omega\Sigma}^0=g_{\omega\Lambda}^0$ and
$g_{\rho\Sigma}^0=2g_{\rho\Xi}^0$. \label{t:t1}}
 \begin{tabular}{ c c c c c c c c c c  c c  }
\hline\hline Model& Case & $g_{\sigma^*\Lambda}$&
$g_{\sigma^*\Sigma}$ & $g_{\sigma^*\Xi}$ & $g_{\sigma\Lambda}^0$
&$g_{\sigma\Sigma}^0$ & $g_{\sigma\Xi}^0$ &
 $g_{\omega\Lambda}^0$  & $g_{\omega\Xi}^0$ & $g_{\rho\Xi}^0$   \\
 SLC & SC  & 6.146&7.651 & 9.764 & 5.920 &3.861 &3.063 &6.884 & 3.442 &3.802 \\
     & UC  & 6.875&10.117 &10.681 & 7.632 &5.573 &7.220 &9.293 & 9.293 &3.802 \\
\hline
SLCd & SC  & 6.146&7.651 & 9.764 & 5.920 &3.861 &3.063 &6.884 & 3.442 &5.776 \\
     & UC  & 6.875&10.117 &10.681 & 7.632 &5.573 &7.220 &9.293 & 9.293 &5.776 \\
\hline\hline
\end{tabular}
\end{center}
\end{table*}

\section{Results and discussions}

Since the parameters for the nucleonic sector of the SLC and SLCd are
clearly given in~\citep{Ji2007b},  we just list in table~\ref{t:t1}
the parameters for the hyperonic sector. We see in table~\ref{t:t1}
that in the same case all parameters but $g_{\rho\Xi}^0$ are the same
for the models SLC and SLCd. This is because the unique difference
between the models SLC and SLCd is that the latter has a softer
symmetry energy than the former. Since the neutron star properties
are rather insensitive to the coupling parameters of $\Sigma$ and
$\Xi$ hyperons owing  to their small fractions in the core of neutron
stars, for simplicity we take $f_{\sigma \Sigma}=f_{\sigma
\Xi}=f_{\sigma \Lambda}$ for the SC in the calculation. For the
similar reason, we prefer the choice $X_{\omega \Lambda}=X_{\omega
\Sigma}=X_{\omega \Xi}$ in the UC calculation, unless otherwise
denoted. For the $\rho$ meson in the UC, the usual relation  $X_{\rho
\Sigma}=2X_{\rho \Xi}=2$ is used. In both the SC and UC, the
$g_{\rho\Lambda}^*$ is zero. Note that all parameters used in this
work for hyperons meet the relation (\ref{eqhpot}).

\begin{figure}
\epsscale{1.0} \plotone{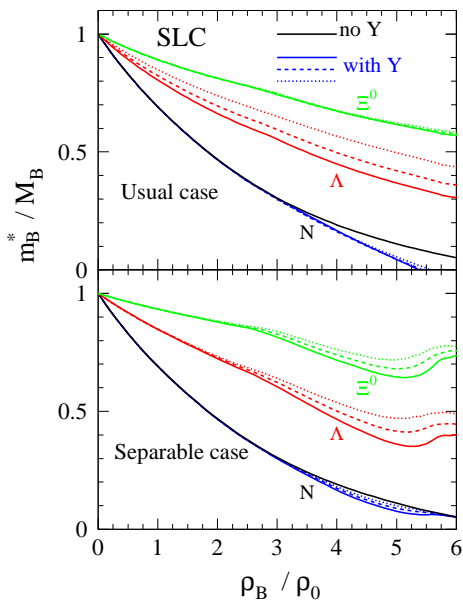} \caption{(Color online) Baryon
effective masses as a function of density in the UC and SC with the
SLC. Three curves (solid, dashed and dotted) of the UC in the upper
panel are calculated with $X_{\omega\Xi}=0.6$ and
$X_{\omega\Lambda}=2/3,0.8$ and 0.9, respectively. In the lower
panel, three curves (solid, dashed and dotted) of the SC are obtained
with $f_{\sigma\Lambda}=f_{\sigma\Xi}= 0.7, 0.8$ and 0.9,
respectively. \label{fmass}}
\end{figure}

As a well-known consequence, the emergence of the hyperon degree of
freedom results in the softening of the EOS. While the models SLC
and SLCd that were constructed based on the BR scaling, the
softening turns out to be too appreciable at high densities to
stabilize the neutron star in the UC with relatively small
$X_{\omega\Lambda}$. As shown in the upper panel of
Fig.~\ref{fmass}, this is related to the rapid decrease of the
nucleon effective mass, corresponding to an increasingly large
scalar field that provides the attraction. As known
before~\citep{Ji2007}, the vector coupling constant is decisive to
generate a stiff EOS at high densities. Thus, the stiffening of the
EOS  can follow from increasing the parameter $X_{\omega\Lambda}$.
With larger $X_{\omega\Lambda}$, for instance,
$X_{\omega\Lambda}=0.9$, the EOS is stiffened to recover the
stability of neutron stars. For the SC, the EOS can be stiffened by
increasing the parameter $f_{\sigma Y}$, since the latter results in
the decrease of the scalar coupling constant, as shown in
Fig.~\ref{fcoup}. Generally, the EOS obtained with the SC is much
stiffer than that with the UC. Meanwhile, the accelerating decrease
of the baryon effective mass due to the inclusion of hyperons can be
greatly suppressed in the SC, as shown in the lower panel of
Fig.~\ref{fmass}. In Fig.~\ref{fmass}, the $\Lambda$ and $\Xi$
hyperon effective masses are also displayed. The upward shift of
hyperon masses at high densities in the lower panel of
Fig.~\ref{fmass} is due to the decrease of the source term (namely,
the hyperon density) of the strange mesons, also see below. The
hyperon effective mass is much larger than the nucleon one. This
would justify the use of the different in-medium interactions for
hyperons and nucleons in the SC.

\begin{figure*}[thb]
\epsscale{1.5} \plotone{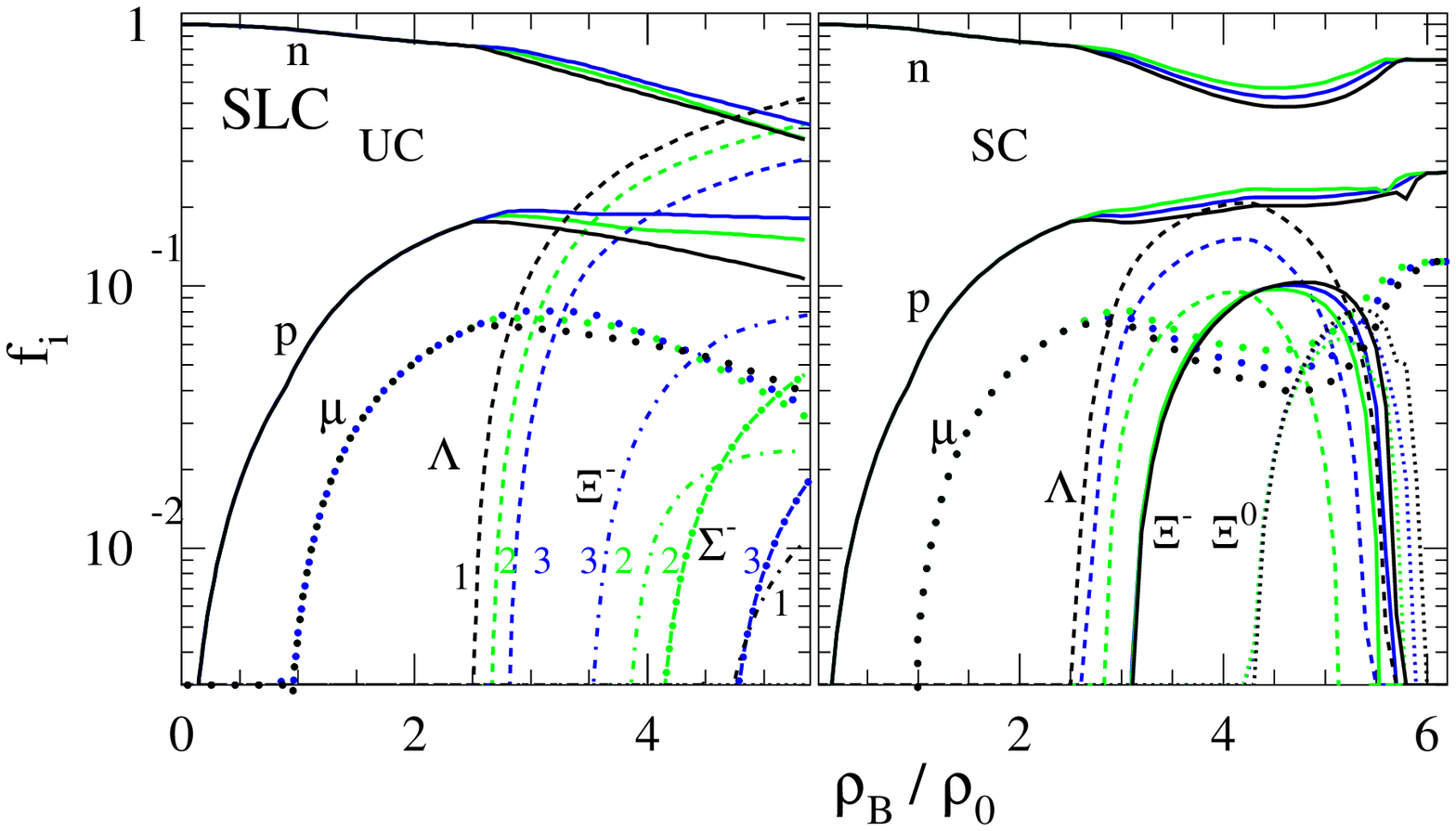} \caption{(Color online) Particle
fractions in the UC and SC  with the model SLC as a function of
density. In the UC (left panel), three curves for each particle,
e.g., denoted by the number 1, 2, and 3, are calculated with
$X_{\omega\Lambda}=2/3$, 0.8 and 0.9, respectively, while
$X_{\omega\Xi}=0.9$. The same number in  the left panel denotes the
results obtained with the same parameters. In the SC (right panel),
three curves for each particle, for instance, in a rising order of
the $\Lambda$ hyperon appearance, are obtained with
$f_{\sigma\Lambda}=f_{\sigma\Xi}= 0.7, 0.8$ and 0.9, respectively.
 \label{ffrac}}
\end{figure*}

The dropping of the baryon  effective mass in chemically equilibrated
matter associates tightly with the beginning density and fractions of
hyperons. In Fig.~\ref{ffrac}, we display the particle fractions as a
function of density. We see that the fractions of hyperons are rather
sensitive to the variation of the in-medium interactions. More
noticeably, quite large differences between the left and right panels
can be observed. For instance, in the UC (left panel), the interval
between the $\Lambda$ and $\Xi^-$ beginning densities depends
sensitively on the ratios of coupling constants, although the
relation (\ref{eqhpot}) is always satisfied, while in the SC such a
sensitivity does not exist. Moreover, the emergence of the $\Sigma$
hyperon depends on the model interaction. The $\Sigma$ hyperon does
not appear in the SC. Remarkably, with the increase of density, the
hyperon fractions in the SC tend downwards till to disappear after
reaching the maximum, as shown in the right panel of
Fig.~\ref{ffrac}. The occurrence of this phenomenon is due to the
fact that the vector meson-hyperon coupling constant $g_{\omega Y}^*$
has a weaker density dependence [see Eq.(\ref{eqphiY})] than the
meson-nucleon coupling constant $g_{\omega N}^*$. At high densities,
$g_{\omega Y}^*$ exceeds $g_{\omega N}^*$, and so do the vector
potentials. The chemical equilibrium thus makes the hyperon Fermi
momenta and fractions lower.  As an application in studying
properties of neutron stars, this actually results in the exclusion
of hyperons in the core of neutron stars and accordingly the
re-stiffening of the EOS at high densities. In addition to our
scheme, we note that there are other attempts to decrease the number
of hyperons in neutron stars. For instance, different couplings of a
new boson to hyperons and nucleons were proposed to decrease the
number of hyperons~\citep{Kr2009}. Including the nonlinear
self-interactions involving a vector meson with hidden strangeness,
Bednarek et al. found that the onset density of hyperons can be as
high as $3\rho_0$ accompanied by smaller fractions of
hyperons~\citep{Be2011}. In~\citep{Ta02,Ts2009}, however, hyperons
were found to appear above $4\rho_0$ and thus played a rather limited
role in the EOS of neutron star matter. In this work, we find that
the onset density of hyperons in the SC can vary upwards within the
region $2.2-3\rho_0$ as the ratio of vector meson coupling
$g_{\omega\Lambda}^{0}/g_{\omega N}^0$ increases from 0.2 to 0.8 (In
table~\ref{t:t1}, this ratio is 2/3), while the hyperon fractions get
suppressed significantly at larger onset densities. A large rise of
this density up to $4.5\rho_0$ can be obtained. However, the binding
relation (\ref{eqhpot}) should then be reduced to
$U^{(N)}_\Lambda(\rho_0)=0$ MeV. In this case, the hyperon fraction
becomes so small that the hyperonic constituents become unimportant.
On the quark level, it is interesting to see that the mechanism of
small numbers of strange quarks in hybrid stars was explored within a
specific quark model~\citep{Bu2004}.

We mention that the baryon fractions in neutron stars are also
sensitive to the symmetry energy. In~\citep{Ji2006}, it was
illustrated that the onset density for hyperons increases moderately
with the softening of the symmetry energy. For the same reason, the
onset density for hyperons with the SLCd is about 0.2$\rho_0$ larger
than those with the SLC. To save space herein, we do not display
particle fractions with the SLCd in a figure similar to
Fig.~\ref{ffrac}.

\begin{figure}[thb]
\epsscale{1.0} \plotone{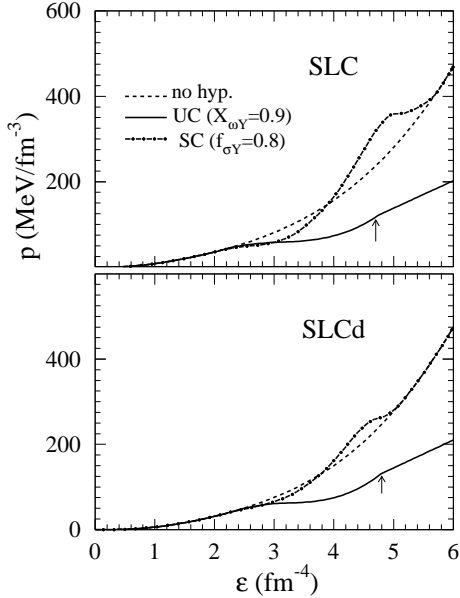} \caption{EOS of
isospin-asymmetric matter with models SLC and SLCd. Curves are
obtained with parameters listed in Table~\ref{t:t1}. The arrow
indicates the critical point to quark matter. The result including
the muon but without hyperons is also depicted for comparisons.
\label{fpeden}}
\end{figure}

Although the emergence of hyperons is a cause for softening the nuclear
EOS, the specific behavior relies indeed on the in-medium
interactions, as clearly shown in Fig.~\ref{fpeden}. The softening
persists at high densities for the UC.  However, the softening is
succeeded by a stiffening in the SC where hyperons feel a
different in-medium interaction from nucleons. Looking back to the
right panel of Fig.~\ref{ffrac}, we see that the stiffening of the
EOS occurs with the suppression of hyperon fractions. As the hyperon
vanishes, the EOS returns to the normal EOS without hyperons. As
shown in Fig.~\ref{fpeden}, the EOS evolves to become stiffer than
the normal one with increasing density. The neutron star matter
thus transits to the normal isospin-asymmetric matter prior to the
vanishing of hyperons. This eventually leaves a limited density
window allowing the existence of hyperons in neutron stars.
Interestingly, we find that the influence of the hyperonization in
the SLCd model falls prominently, as compared to the SLC model. This
is attributed to the larger onset density of hyperons with the
SLCd due to the softening of the symmetry energy, as mentioned above.

In the UC, since the nucleon effective mass vanishes at certain
critical density, we need to consider the EOS beyond the critical
density. Though the relationship between the chiral restoration and
deconfinement occurrence is still discussed, it is nevertheless a
convenient and usual way to neglect the distinction between them.
Beyond the critical density, we thus adopt a quark matter EOS
described by the MIT model with the appropriate bag parameter
[$(179.5 MeV)^4$] to connect smoothly the hadron matter and quark
matter EOSs. The connection to the quark matter further softens the
EOS, as shown in Fig.~\ref{fpeden}.

\begin{figure}[thb]
\epsscale{1.0} \plotone{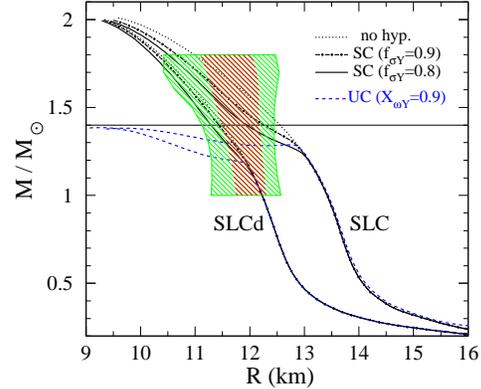} \caption{(Color online)
Mass-radius relation of neutron stars in the UC and SC with the SLC
and SLCd. In the SC, two curves are obtained with $f_{\sigma Y}=0.8$
and 0.9, respectively. The hatched areas give the probability
distributions with $1\sigma$ (red) and $2\sigma$ (green) confidence
limits for $r_{ph}\gg R$ summarized in~\citep{St2010}. \label{fmsr}}
\end{figure}

We now turn to the consequences of hyperonizations on properties of
neutron stars. In particular, it is interesting to see whether our
model with hyperonization can give properties of neutron stars that
are compatible with the recent observation of the millisecond pulsar
J1614-2230. The mass of this pulsar was accurately determined to be
$1.976\pm0.04 M_\odot$~\citep{De10}. It was concluded that such a
large mass can rule out almost all currently proposed hyperon
equations of state. The mass-radius relation of neutron stars is
obtained from solving the standard TOV equation with the nuclear EOS
being specified above. For the low-density crust, we adopt the EOS
in~\citep{Ba1971,Ii1997}.  It is seen in Fig.~\ref{fmsr} that all
EOS's of the SC can produce neutron stars with the maximum mass
around $2M_\odot$ and with the corresponding radius ranging from
$9.3$ to $9.5$ km. Since the maximum mass of neutron stars is more
sensitive to the EOS at high densities, it is understandable that
the EOS including hyperonization at intermediate densities can lead
to masses compatible with that of the pulsar J1614-2230. Thus, the
EOS with hyperonization can not be simply ruled out. Of course, the
hyperon fraction in neutron stars with the SC is rather limited, as
compared to that with the UC. In addition, a subtle factor affecting
the hyperonization is the density dependence of nuclear symmetry
energy. Due to the softening of the symmetry energy in the SLCd, the
hyperonization in the SC becomes unimportant for the EOS of neutron
star matter, see Fig.~\ref{fpeden}, which equivalently expels most
hyperons in neutron stars. Consequently, the mass-radius relation
with the SLCd is almost independent of the parameter $f_{\sigma Y}$
in the SC, as shown in Fig.~\ref{fmsr}.  For the UC, it looks
undoubtedly that the EOS is ruled out by the recent
observation~\citep{De10}, since the maximum mass of neutron stars in
this case just sprints to $1.4M_\odot$ with a much compacter size
than the canonical neutron star without hyperons.

It is worth adding here some discussions about the influence of the
symmetry energy on the radii of neutron stars. It is now well
established that the maximum mass of neutron stars is dominated by
the high-density behavior of the EOS, while the radius is primarily
determined by the slope of the symmetry energy at intermediate
densities (1 -3$\rho_0$) ~\citep{La2001,La2004,St2005,Xu2009}. The
large difference between the radii of low-mass neutron stars obtained
with the SLC and SLCd can be attributed to the difference in the
slope parameter $L$. As discussed in detail earlier in ref.~\citep{Fat10},
the central density of an 18 - 20 km star is near $\rho_0$, and the crust
ends approximately at $(1/3-1/2)\rho_0$. In this density range, the pressure
is dominated by the symmetry energy and not the incompressibility of
symmetric nuclear matter. Our results shown in Fig. 5 are consistent
with those in ref.~\citep{Fat10}. For the 1.4$M_\odot$ neutron stars in the SC, we
see however that the inclusion of hyperons reduces moderately the
range of the neutron star radius. For instance, the radius ranges
from 11.4km with the SLCd to 12.3km with the SLC for $f_{\sigma
Y}=0.9$, which are well situated in the domain extracted by Steiner
et al.~\citep{St2010}. With the emergence of hyperons, the
sensitivity of the star radius to the symmetry energy is reduced
clearly. This is because the $\Lambda$-hyperon, being the dominant
component of hyperons, is an isospin scalar. The suppression of the
isovector potential $g_\rho^*b_0$  due to the appearance of $\Lambda$
hyperons~\citep{Ji2006} results in the reduction of the pressure of
asymmetric matter and thus the star radius, while the magnitude of
the suppression depends on the specific values of the symmetry energy
and hyperon fraction. As a result, the less sensitivity of the star
radius to differences in the symmetry energy is observed in
calculations with both the SC and UC. In Fig.~\ref{fmsr}, we also
include the constraints of mass-radius trajectories for $r_{ph}\gg R$
obtained by Steiner et al. in~\citep{St2010}. Our results in the SC
are either within (for SLCd) or not very far off (for SLC) the
constrained region, though the inclusion of hyperons seems to tilt
the vertical trajectories. For low-mass neutron stars, the radii with
the SLC are predicted to go beyond the optimal region extracted very
recently in~\citep{St2012b}, unless some other scenarios that allow a
loose extension of the radii are invoked to extract the radius
constraints~\citep{Su2011,Zh2007}. However, the maximum mass of
neutron stars is almost independent of the slope $L$ and the radii of
low-mass neutron stars. In our cases, the maximum mass is only
reduced by about 1\% by softening the symmetry energy from the model
SLC to SLCd. On the other hand, it is known that the measurements of
the neutron star radii are far less precise than the mass
measurements, see, e.g., refs.~\citep{Su2011,Zh2007} and references
therein. We note that an option of $r_{ph}= R$ can cause a visible
slanting of the vertical trajectories~\citep{St2010}. The agreement
of our results can thus be better with the constraints obtained for
$r_{pc}=R$.

We note that there had been a few endeavors in the past to involve
the hyperons in heavy neutron stars either by invoking strong
repulsions for hyperons or pushing upwards the onset density of
hyperons\citep{Ho01,Ta02,St07,De08,Be2011,We11}. The main purpose of
our work is to constrain the in-medium hyperon potentials using
the 2 solar mass constraint of neutron stars. This indeed requires
strong repulsions for hyperons. Due to different density-dependencies
of nucleonic and hyperonic potentials in the SC, the hyperonic vector
potential exceeds the nucleonic one, leading to a very significant
suppression of hyperons at high densities. As a result, the hyperons
would exist in a shell of neutron star core even with a small $L$
value.

Besides the effect of hyperonization on the static properties,
another consequence of the hyperonization is on the thermal evolution
of neutron stars. For the SC in SLC, we see from Fig.~\ref{fmsr} that
$\Lambda$ hyperons start to appear above $2.5\rho_0$. At this central
density the neutron star mass is about 1.2$M_\odot$. For most
observed neutron stars that have larger masses, it appears that the
direct Urca (DU) process with nucleons \citep{La1991} and/or hyperons
\citep{Pr1992} would occur since the proton fraction in the neutron
star matter with hyperonization can be in excess of the DU threshold
(14\%) with the SC.   According to the thermal evolution of observed
neutron stars analyzed in \citep{Pa2004,Ya2004}, the fast cooling
with the DU processes seemed to be excluded in most neutron stars
except the massive ones. Slower cooling is possible when the neutrino
emissivity can be suppressed by the superfluity of constituent
particles like nucleons and hyperons when the temperature falls below
the critical temperature. Page et al. set a stringent requirement for
the critical temperature of neutron superfluidity without which
enhanced cooling from DU processes may be needed in at least half of
the observed young cooling neutron stars~\citep{Pa2009}. While the DU
cooling involving nucleons only was regarded to be too fast, it was
pointed out by Tsuruta et al. that the DU cooling with hyperons in
neutron stars can be compatible with the observations, provided that
the hyperon superfluidity is appropriately accounted
for~\citep{Ts2009}. In our models with the SC, the hyperonization can
thus be favorable to make up  the potential incompatibility in the
thermal evolution by considering the hyperon superfluidity. On the
other hand, the threshold mass of neutron stars which allow the DU
process increases with the softening of the symmetry energy due to
which  the proton fraction exceeds the threshold value at larger
densities. For instance, this mass is around 1.3$M_\odot$ with the SC
of the model SLCd, and the threshold density for the DU process is
about 4$\rho_0$, which is coincidently the onset density of hyperons
in~\citep{Ts2009}.

Finally, we discuss some details concerning the in-medium
interactions for hyperons. Constrained by the large mass of observed
pulsars, it is favorable for us to select the in-medium interactions
for hyperons in the SC. It is however interesting to find that the
single-particle potentials for hyperons in the UC and SC almost
overlap in a large density domain ranging from zero density to
intermediate density. The significant departure appears only at high
densities (roughly $\ge 4\rho_0$). Especially at lower densities, the
SC and UC descriptions give rather limited differences in
single-particle properties. This thus indicates that without
compromising the success in describing properties of hypernuclei, one
can constrain significantly the high-density hyperon interactions
with the large mass of neutron stars. In addition, we have noticed that
many efforts using various quark models with the postulate of
strong interactions and/or color superconductivity can also lead to
large-mass hybrid stars~\citep{Al05,Kl2007,Ip2008,We11b,Bo2012} with
stiff EOS's at high densities. Our models respecting chiral limits
possess the similarly stiff EOS at high densities. On the other hand,
neutron stars may be composed of more complicated constituents
including quarks and meson condensates. It is useful to consider
these non-baryonic degrees of freedom and their interplay with
hyperons to constrain more quantitatively the in-medium interactions
for hyperons. However, this is beyond the scope of the present work.

\section{Summary}
The density-dependent relativistic mean-field model, which was
constructed to respect the chiral symmetry restoration at high
densities in terms of in-medium hadron properties according to the
Brown-Rho scaling, is extended to include the hyperonization in
isospin-asymmetric matter. We examined effects of the in-medium
hyperon interactions on properties of neutron stars. It is found
that the maximum mass of neutron stars can constrain significantly
the in-medium hyperon interactions. In particular, assuming two
categories of in-medium interactions for hyperons, we investigated
their distinct roles in hyperonizations and properties of neutron
stars. With different in-medium interactions for hyperons and
nucleons (i.e., the SC case), a maximum neutron star mass of
$2M_\odot$ can be obtained with the model where the nucleonic EOS is
consistent with the terrestrial nuclear laboratory data. The result
in the SC is compatible with recent observation on the mass of the
millisecond pulsar J1614-2230. In this scheme, the number of
hyperons is limited in neutron stars. Interestingly, we also found
that the softening of the symmetry energy can play an important role
in further reducing hyperons in neutron stars. On the contrary, the
scheme that adopts similar in-medium interactions for both hyperons
and nucleons (i.e., the UC case) can give a maximum neutron star
mass of only about $1.4M_\odot$. Nevertheless, the hyperonization in
both cases reduces clearly the sensitivity of the neutron star
radius to the difference in the nuclear symmetry energy.

\section*{Acknowledgement}
We would like to thank William Newton for his very helpful and
constructive comments. One of the authors (J.W.Z.) would like to
thank Ang Li and Zhao-Qing Feng for useful discussions and expresses
his sincere gratitude to Dr. Song Dan at the affiliated Hospital of
Nanjing Medical University for his great help during the time of
this work. This work was supported in part by the National Natural
Science Foundation of China under Grant Nos. 10975033, 10975097 and
11135011, the China Jiangsu Provincial Natural Science Foundation
under Grant No.BK2009261, the China Major State Basic Research
Development Program under Contract No. 2007CB815004, Shanghai
Rising-Star Program under Grant No.11QH1401100, the ``Shu Guang"
project supported by Shanghai Municipal Education Commission and
Shanghai Education Development Foundation, the Program for Professor
of Special Appointment (Eastern Scholar) at Shanghai Institutions of
Higher Learning, the Science and Technology Commission of Shanghai
Municipality (11DZ2260700), the US NSF under grants PHY-0757839 and
PHY-1068022 and NASA under grant NNX11AC41G issued through the
Science Mission Directorate.

\bibliographystyle{yahapj}

\end{document}